\documentclass[12pt]{article}

\usepackage{verbatim, color, graphics}

\usepackage{graphicx}
\usepackage{epsfig}
\usepackage{amssymb}
\usepackage{amsthm}
\usepackage{bm}
\usepackage{amsmath}
\usepackage{color}

\begin{document}

\title{The effect of an exterior electric field on the instability of dielectric plates}

\author{
Yipin Su$^{1,2}$, Weiqiu Chen$^{1}$, Luis Dorfmann$^{3}$ and Michel Destrade$^{2,1}$
\\[12pt]
$^{1}$Department of Engineering Mechanics, \\
Zhejiang University, Hangzhou 310027, P.R. China\\
$^{2}$Stokes Centre for Applied Mathematics, \\
School of Mathematics, Statistics and Applied Mathematics, \\
NUI Galway, University Road, Galway, Ireland\\
$^{3}$Department of Civil and Environmental Engineering, \\
Tufts University, Medford, MA 02155, USA}

\date{}


\maketitle

\begin{abstract}
\color {black} We investigate the theoretical nonlinear response, Hessian stability, and possible wrinkling behaviour of a voltage-activated dielectric plate immersed in a tank filled with silicone oil.  \color{black} 
Fixed rigid electrodes are placed on the top and bottom of the tank, and an electric field is generated by a potential difference between the electrodes. 
We solve the associated incremental boundary value problem of superimposed, inhomogeneous small-amplitude wrinkles, signalling the onset of instability. 
We decouple the resulting bifurcation equation into symmetric and antisymmetric modes. 
\color{black} 
For a neo-Hookean dielectric plate, we show that  a potential difference between the electrodes can induce a thinning of the plate and thus an increase of its planar area, similar to the scenarios encountered when there is no silicone oil.
However, we also find that, depending on the material and geometric parameters, an increasing applied voltage can also lead to a \textit{thickening} of the plate, and thus a shrinking of its area. 
In that scenario, Hessian instability and wrinkling bifurcation may then occur spontaneously once some critical voltages are reached.
 \color{black} 
\end{abstract}


\section{Introduction}


Dielectric elastomers (DEs) are soft  materials capable of undergoing large deformations when activated by an  electric field.  Compared with other `smart' materials such as, for example, electro-active ceramics and shape memory alloys, DEs have the advantages of high-sensitivity, low noise and large actuation strains.
They have potential applications as sensors,  as artificial muscles and as biologically inspired intelligent devices \cite{Bar-Cohen04, Brochu10, Li2018}.

Typically, a planar dielectric actuator consists of a thin electro-elastic plate with two  electrodes coated on its main surfaces. 
Carbon grease or a conductive hydrogel are frequently used as  electrodes.  
These coated films are  thin and highly flexible and  do not constrain the deformation of the actuator,  hence play no noticeable mechanical role. 
Applying a potential difference  \cite{ZhSu07, Koh2011, Che17}  or a  surface charge density  \cite{Keplinger2010, Li2011} on the electrodes generates electrostatic  forces causing a reduction in thickness and therefore, by incompressibility,  a planar  expansion. 
When the  thickness of the plate is small compared to the planar dimensions, there is  no electric field outside the plate by Gauss's theorem.

Another possible  actuation is obtained by immersing a dielectric plate in a conductive fluid, with the electric field generated by a potential difference  or by a surface charge density on two rigid electrodes positioned appropriately.  Notable applications of that setup include biometric sensors and wearable or implantable electronic devices. 
For example, Sun et al. \cite{Sun2014} designed an `ionic skin' consisting of a stretchable dielectric membrane sandwiched between two flexible  ionic conductors using rigid electrodes. 
Their device operates as a  strain sensor and can be used to  detect the location and pressure of human touch.  Liquid crystal elastomers and ionic electro-active polymers are considered electro-active elastomers with potential applications in display technologies and drug delivery systems \cite{O'Halloran08}. 
Bioelectrical and mechanical interactions exist in biological tissues such as bones, vessels and nerve cells \cite{Fukada1957, Fukada1968} and,  more recently,  have been identified as powerful tools of biological pattern control \cite{Pietak2016}.   
For an electro-active elastomer immersed in a conductive medium with non-zero permittivity, the surrounding electric field has an obvious and significant influence on the mechanical response of the plate.

Advanced theories to describe the main electro-mechanical phenomena  observed in DEs have been developed over the years, including  nonlinear deformation \cite{Dorf14b, Bortot17}, vibration \cite{Zhu2010, Wang2016, Jin2017}, wrinkling instability \cite{Galich17, Yang17, Su2018} and wave propagation \cite{Shmuel2013, Wu2017, Drofmann2020}.  
Most of these studies focused on electro-elastic plates  actuated by flexible electrodes. 
In contrast, little attention has been devoted to devices actuated by an external  electric field. 

Dorfmann and Ogden \cite{Dorf10} studied the large plane strain deformation and instability of an electro-elastic  half-space in the presence of a uniform external electric field normal to its surface. 
Subsequently, they studied the instability of an equi-biaxially deformed dielectric plate, without flexible electrodes, but in the presence of an external electric field normal to its faces \cite{Dorf14a}.  
Chen and Dai \cite{Chen2012} investigated the propagation of axisymmetric waves in a homogeneously deformed dielectric cylinder subject to an external electric field oriented along the axial direction and derived the exact wave solution in terms of Bessel functions.  
Su et al. \cite{Su16a, Su2016b} examined the propagation of non-axisymmetric waves in, and the instability of, an electro-active hollow cylinder under uniform mechanical and electrical fields.  
They found that the exterior electric field can have a stabilising or destabilising influence, depending on the electro-elastic coupling parameters of the materials.  Note that in these studies the DEs were surrounded by vacuum to simplify the forms of the governing equations and of the boundary conditions.

D\'iaz-Calleja et al. \cite{Diaz-Calleja2009} compared the pull-in instabilities of an electro-elastic  plate with two flexible electrodes coated on its faces and of a dielectric plate floating between two fixed rigid electrodes.  
They investigated the effect of the permittivity of the surrounding fluid on the stability of the elastomer. 
They focused on homogeneous instabilities only and did not include inhomogeneous wrinkling-type \color{black}modes\color{black}. 
\color{black} Wang et al. \cite{Wang2011}  observed experimentally that instabilities occur in a substrate-bonded elastomer film immersed in a conductive solution sandwiched between electrodes when the applied voltage exceeds a critical value. 
They predicted the critical field for instability by comparing the potential energies in the wrinkled and flat states. However, they did not consider the mechanical role of the conductive solution.\color{black}

\begin{figure}[t!]
\centering
\includegraphics[angle=-90,width=0.9\textwidth]{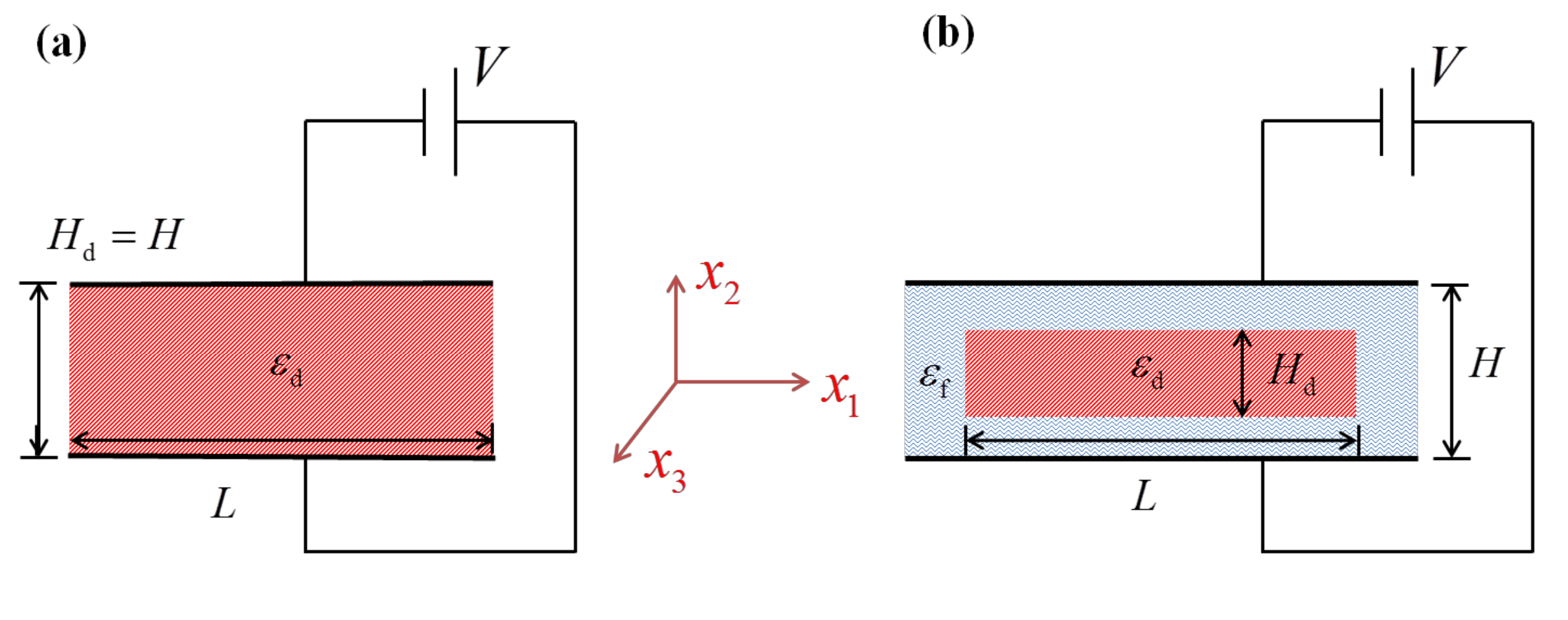}
\caption{
{\footnotesize
Loading scenarios for an electro-elastic plate. In (a) the electric field is generated by a potential difference between two flexible electrodes coated on the main faces, in (b) the plate is immersed in a tank filled with silicone oil with the electric field generated by two  rigid electrodes placed at the top and bottom of the tank.}
}
\label{figure1}
\end{figure}

\color{black}
Su et al. \cite{Su2018} studied the nonlinear deformation and wrinkling behaviour of a dielectric plate sandwiched between two compliant electrodes (Figure \ref{figure1}a). 
They found that a superposed mechanical load is required to induce wrinkling of the plate. 
In this paper, we consider a DE plate immersed in silicone oil (Figure \ref{figure1}b).
In particular, we show that the plate can lose stability by applying a voltage only, due to electrostriction.\color{black} 

Consider a rectangular electro-elastic plate with permittivity $\varepsilon_{\mathrm d}$ and with reference  thickness $H_{\mathrm d}$.  
Following the application of an electric field, the resulting deformation is homogeneous and defined in terms of the principal  stretches   $\lambda_1,\lambda_2,\lambda_3$. 
The material is incompressible, so that the change in thickness is defined by the stretch $\lambda_2=\lambda_1^{-1}\lambda_3^{-1}$. 
Hence, the deformed plate has uniform thickness

\begin{equation}
h_{\mathrm d} =\lambda_2 H_{\mathrm d}=\lambda_1^{-1}\lambda_3^{-1}H_{\mathrm d}.
\end{equation} 
 
\color{black}
For a DE plate immersed in a tank filled with silicone oil having an electric permittivity $\varepsilon_{\mathrm f}$, the electric field is generated by a potential difference, say $V$, between two fixed rigid electrodes placed on the top and bottom of the tank.  
The distance between the two electrodes is constant and denoted $H$. 
It follows that
\begin{equation}
\label{electric-field-oil}
E_0 H_{\mathrm d}+E^{\star}(H-h_{\mathrm d})=V,
\end{equation}
where $E_0$ is the nominal electric field in the plate and $E^{\star}$ denotes the electric field in the silicone oil.  
Throughout the paper we use the star superscript $^{\star}$ to denote quantities defined in the oil, outside the region occupied by the plate.
\color{black}

The  paper is organised as follows. 
In  Section \ref{section2} we present the  equations governing the equilibrium of an electro-elastic plate and specialise the boundary conditions to the setup  considered here (Figure \ref{figure1}b).  

Then, in Section \ref{section3}, we \color{black} extend the method of Su et al. \cite{Su2018} to include the effect of an exterior electric field. \color{black}  
\color{black}
Hence we consider a 2D incremental displacement superposed upon the large equi-biaxial deformation. 
\color{black} Specifically, we look for small sinusoidal variations in the $x_1-$direction with exponential variation in the $x_2-$direction and  rewrite the incremental equations in the Stroh form to solve the corresponding boundary value problem.  
We manage to factorise the resulting \color{black}wrinkling \color{black}  equation of the plate into symmetric and antisymmetric modes.  

In Section \ref{section4} we consider the equilibrium of an equi-biaxially deformed electro-elastic plate and its  stability, when the material is modelled by the neo-Hookean dielectric model.  
We derive explicitly the bifurcation \color{black} criterion \color{black} for neo-Hookean  electro-elastic plates of finite thickness and its specialisations in the \color{black} thin-plate and thick-plate \color{black} limits.  
Our numerical results elucidate the influences of the material properties of the  silicone oil and of the elastomer on the nonlinear response and on the instability behaviour of the electro-elastic plate.
In particular we find scenarios where the immersed plate \color{black} (Figure \ref{figure1}b) \color{black} can be made to wrinkle by voltage alone, a possibility that does not arise when there is no exterior field \color{black} (Figure \ref{figure1}a)\color{black}. 
We give concluding remarks in Section \ref{section5}.


\section{Large actuation}
\label{section2}


According to the nonlinear theory of electro-elasticity \cite{Dorf06, Suo08, Dorfmann2017}, for an incompressible dielectric plate in mechanical and electrostatic equilibrium, we find that
\begin{align}
\label{constitutive}
\tau_{11}-\tau_{22}=\lambda_1 \frac{\partial W}{\partial \lambda_1},\qquad
\tau_{33}-\tau_{22}=\lambda_3 \frac{\partial W}{\partial \lambda_3}, \qquad
D_0=-\frac{\partial W}{\partial E_0},
\end{align}
where $\tau_{ii}\ (i=1,2,3)$ is the principal component of the total Cauchy stress tensor of the solid in the $x_i-$direction, $W(\lambda_1,\lambda_3,E_0)$ is the energy density of the elastomer and $D_0$ is the non-zero component of the nominal electric displacement field in the $x_2-$ direction inside the plate. 

%
%
%
%

The true electric displacement $D$ \color{black} in the plate \color{black} is connected to its nominal counterpart $D_0$  by
\begin{equation}
D=\lambda_1^{-1}\lambda_3^{-1} D_0=-\lambda_1^{-1}\lambda_3^{-1}\frac{\partial W}{\partial E_0}.
\end{equation}

The normal component of the electric displacement, in the absence of free surface charge,  is continuous across a material interface. It follows that 
\begin{equation}
\label{exterior-electric-displacement}
D^{\star}=D=-\lambda_1^{-1}\lambda_3^{-1}\frac{\partial W}{\partial E_0},
\end{equation}
where $D^{\star}$ denotes the electric displacement in the silicone oil in the $x_2-$direction.  The electric field in the silicone oil can now be determined as
\begin{equation}
E^{\star}=\varepsilon_{\mathrm f}^{-1} D^{\star}=-\varepsilon_{\mathrm f}^{-1}\lambda_1^{-1}\lambda_3^{-1}\frac{\partial W}{\partial E_0},
\end{equation}
so that Eq. \eqref{electric-field-oil} reads
\begin{equation}
\label{electric-field}
E_0 H_{\mathrm d}-\varepsilon_{\mathrm f}^{-1}\lambda_1^{-1}\lambda_3^{-1}\frac{\partial W}{\partial E_0}(H-\lambda_1^{-1}\lambda_3^{-1}H_{\mathrm d})=V.
\end{equation}
For a given energy function $W(\lambda_1,\lambda_3,E_0)$,  the nominal electric field in the plate $E_0$ is then obtained from Eq. \eqref{electric-field}.

The non-zero components of the Maxwell stress tensor in the silicone oil have the forms \cite{Dorf14a}  
\begin{equation}
\label{exterior-stressa}
\tau_{11}^{\star}=\tau_{33}^{\star}=-\tau_{22}^{\star}=-\frac{1}{2\varepsilon_{\mathrm f}}\left(D^{\star}\right)^2=
-\frac{1}{2\varepsilon_{\mathrm f}}\lambda_1^{-2}\lambda_3^{-2}\left(\frac{\partial W}{\partial E_0}\right)^2.
\end{equation}

\color{black} Assume that the plate is subject to in-plane biaxial nominal stresses $s_1$ and $s_3$ in the $x_1-$ and $x_3-$directions, respectively, with its main surfaces $x_2=\pm h_{\mathrm d}/2$ being traction-free. \color{black} 
The mechanical continuity condition gives 
\begin{equation}
\label{general-boundary2}
s_1=\lambda_1^{-1}\left(\tau_{11}-\tau_{11}^{\star}\right), \qquad s_3=\lambda_3^{-1}\left(\tau_{33}-\tau_{33}^{\star}\right), \qquad \tau_{22}=\tau_{22}^{\star}.
\end{equation}

Finally, combining Eqs.~\eqref{constitutive} and \eqref{general-boundary2} yields the  equations governing the nonlinear response of the electro-elastic plate:
\begin{align}\label{problem2}
\frac{\partial W}{\partial \lambda_1}+\lambda_1^{-1}\left(\tau_{22}^{\star}-\tau_{11}^{\star}\right)=s_1,\qquad \frac{\partial W}{\partial \lambda_3}+\lambda_3^{-1}\left(\tau_{22}^{\star}-\tau_{33}^{\star}\right)=s_3.
\end{align}


\section{Linearised stability analysis}
\label{section3}


\subsection{Stroh formulation}

In this section we use the  linearised incremental theory of electro-elasticity to predict the onset of wrinkling instability for the actuated electro-elastic plate  \cite{Dorf10, Dorfmann2019,  Bertoldi2011, Su2016b, Rudykh2017, Shmuel2018}.  
Here and in the following, we use a superimposed dot to denote  increments. 

Superimpose a small-amplitude mechanical perturbation $\mathbf u=\bold{\dot x}$ along with an updated incremental electric field $\bold{\dot {\mathbf E}}$ upon the finitely deformed configuration. 
The corresponding updated incremental electric displacement field is $\bold{\dot{\mathbf D}}$. 
We look for incremental two-dimensional solutions  independent of $x_3$. 
A simple analysis then shows that $u_3=\dot{E}_{3}=\dot{D}_{3}=0$.

The incremental electric field is irrotational and, therefore, it can be expressed as
\begin{equation}
\label{electricfield}
\dot{E}_{1}=-\frac{\partial\dot{\phi}}{\partial x_1},\qquad
\dot{E}_{2}=-\frac{\partial\dot{\phi}}{\partial x_2},
\end{equation}
where $\dot{\phi}$ is the incremental electric potential in the plate.

Following Su et al. \cite{Su2018}, we seek solutions of the incremental equilibrium equations  that vary sinusoidally along the $x_1-$direction with amplitude variations in  the $x_2-$direction. 
We use the form
\begin{equation}
\label{incremental-fields}
\left\{ {{u}_{1}},{{u}_{2}},{{{\dot{D}}}_{2} ,{{{\dot{T}}}_{21}},{{{\dot{T}}}_{22}}},\dot{\phi} \right\}=  
\Re \left\{\left[k^{-1} U_1, k^{-1} U_2,\textrm{i} \Delta,\textrm{i} \Sigma_{21},\textrm{i} \Sigma_{22}, k^{-1} \Phi\right] e^{\textrm{i} kx_1}\right\},
\end{equation}
where $\dot T_{21}, \dot T_{22}$ are the components of the updated incremental nominal stress, $U_1$, $U_2$, ${{\Delta }}$, ${{\Sigma }_{21}}$, ${{\Sigma }_{22}}$, $\Phi$ are functions of $k{{x}_{2}}$ only, and $k=2\pi/\mathcal L$ is the wavenumber with $\mathcal L$ being the wavelength of the wrinkles of the buckled plate.

It is convenient  to rewrite the  boundary value problem in the Stroh form,
\begin{equation}
\label{stroh}
\boldsymbol{\eta}'=\textrm{i}\mathbf{N}\boldsymbol{\eta}=\textrm{i}\left[ \begin{matrix}
{{\mathbf{N}}_{1}} & {{\mathbf{N}}_{2}}  \\
{{\mathbf{N}}_{3}} & \mathbf{N}_{1}^T \\
\end{matrix} \right]\boldsymbol{\eta},
\end{equation}
where $\mathrm i^2=-1$, $\mathbf N$ is the Stroh matrix, 
$\boldsymbol{\eta} ={{\left[ \begin{matrix}
U_1 & U_2 & \Delta & \Sigma_{21} & \Sigma_{22} & \Phi  \\
\end{matrix} \right]}^{T}}$ 
is the Stroh vector \cite{Destrade09, Su2018}, and the prime  denotes differentiation with respect to $k{{x}_{2}}$. 
In what follows we use the  notation $\boldsymbol\eta=\left[\begin{matrix}\mathbf U & \mathbf S\end{matrix} \right]^T $, where   $\mathbf U={{\left[ \begin{matrix}
   U_1 & U_2 & \Delta \\
\end{matrix} \right]}^{T}}$ and $\mathbf S={{\left[ \begin{matrix}
  \Sigma_{21} & \Sigma_{22} & \Phi  \\
\end{matrix} \right]}^{T}}$  represent  the generalised displacement and traction vectors, respectively.
Here we find that the $3\times 3$ sub-blocks ${{\mathbf{N}}_{1}}$, ${{\mathbf{N}}_{2}}$ and ${{\mathbf{N}}_{3}}$ have the  components
\begin{align}\label{stroh-components}
& \mathbf{N}_1 = \left[ \begin{array}{ccc}
0 & -1+\tau_{22}/c & 0 \\
-1 & 0 & 0 \\
0 & d\tau _{22} /c & 0
\end{array} \right],
\qquad
\mathbf{N}_2 = \left[ \begin{array}{ccc}
1/c & 0 & d/c \\
0 & 0 & 0 \\
d/c & 0 & d^2/c -f
\end{array} \right],
\nonumber \\[12pt]
&
\mathbf{N}_3 = \left[ \begin{array}{ccc}
-2(b+c-\tau_{22})+e^2/g & 0 & -e/g \\
0 & -a + (c-\tau _{22})^2/c &0 \\
-e/g & 0 & 1/g 
\end{array} \right],
\end{align}
where $a$, $b$, $c$, $d$, $e$, $f$, $g$ are electro-elastic coefficients evaluated for a prescribed finite deformation and for a given material model, see \cite{Su2018} for their general expressions.

Because the Stroh matrix $\mathbf N$ is constant, the solution to Eq. ~\eqref{stroh} is of the exponential form,
\begin{equation}
\boldsymbol{\eta}(kx_2) = \left[ \begin{array}{c} \mathbf{U}(kx_2) \\
\mathbf{S}(kx_2) \end{array} \right] = \sum _{j=1} ^6 c_j \boldsymbol{\eta} ^{(j)} e^{\mathrm i q_j k x_2}, \label{eta_plate}
\end{equation}
where   $c_j, \ j=1,2,\dots,6,$ are  arbitrary constants to be determined and   $q_j,\ j=1,\dots, 6,$ are the eigenvalues found by solving the  characteristic equation
\begin{equation}
\label{eigensystem}
\det \left(\mathbf N-q\mathbf I\right)=0.
\end{equation}
The eigenvectors  $\boldsymbol{\eta}^{(j)}, \ j=1,2,...,6$ associated with $q_j$ \color{black} are connected through the relation $\boldsymbol{\eta}^{(j)}= ({\boldsymbol{\eta}}^{(j+3)})^*$, $j=1,2,3$, where the asterisk denotes the complex conjugate. \color{black}
Note that for the model used in the next section (neo-Hookean model) we  have the connections  $q_j=\mathrm{i} p_j$ and $q_{j+3}=-\mathrm{i} p_j$, $j=1,2,3$ with real $p_j$.

We now write the generalised displacement and traction vectors at the faces $x_2=\pm h_{\mathrm d}/2$ in the  forms
{\small
\begin{equation}
\label{generalised-displacement}
\left[ \begin{array}{c} \mathbf{U}(kh_{\mathrm d}/2) \\
\mathbf{U}(-kh_{\mathrm d}/2) \end{array} \right] = \left[ \begin{array}{cccccc} 
\eta^{(1)}_1 E_1 ^- & \eta^{(2)}_1 E_2 ^- & \eta^{(3)}_1 E_3 ^- & \eta^{(4)}_1 E_1 ^+ & \eta^{(5)}_1E_2 ^+& \eta^{(6)}_1 E_3 ^+\\
\eta^{(1)}_2 E_1 ^- & \eta^{(2)}_2E_2 ^-& \eta^{(3)}_2E_3 ^- & \eta^{(4)}_2E_1 ^+ & \eta^{(5)}_2E_2 ^+ & \eta^{(6)}_2E_3 ^+ \\
\eta^{(1)}_3E_1 ^- & \eta^{(2)}_3E_2 ^-& \eta^{(3)}_3E_3 ^-& \eta^{(4)}_3E_1 ^+ & \eta^{(5)}_3E_2 ^+ & \eta^{(6)}_3E_3 ^+ \\
\eta^{(1)}_1E_1 ^+ & \eta^{(2)}_1E_2 ^+ & \eta^{(3)}_1E_3 ^+ & \eta^{(4)}_1E_1 ^- & \eta^{(5)}_1E_2 ^- & \eta^{(6)}_1E_3 ^- \\
\eta^{(1)}_2E_1 ^+ & \eta^{(2)}_2E_2 ^+ & \eta^{(3)}_2E_3 ^+ & \eta^{(4)}_2E_1 ^- & \eta^{(5)}_2E_2 ^- & \eta^{(6)}_2E_3 ^- \\
\eta^{(1)}_3E_1 ^+ & \eta^{(2)}_3E_2 ^+ & \eta^{(3)}_3E_3 ^+ & \eta^{(4)}_3E_1 ^- & \eta^{(5)}_3E_2 ^- & \eta^{(6)}_3E_3 ^-  \end{array} \right] \left[ \begin{array}{c} c_1 \\ c_2 \\ c_3 \\ c_4 \\ c_5 \\ c_6 \end{array} \right], 
\end{equation}

\begin{equation} 
\label{generalised-stress}
\left[ \begin{array}{c} \mathbf{S}(kh_{\mathrm d}/2) \\
\mathbf{S}(-kh_{\mathrm d}/2) \end{array} \right] = \left[ \begin{array}{cccccc} 
\eta^{(1)}_4 E_1 ^- & \eta^{(2)}_4 E_2 ^- & \eta^{(3)}_4 E_3 ^- & \eta^{(4)}_4 E_1 ^+ & \eta^{(5)}_4 E_2 ^+ & \eta^{(6)}_4 E_3 ^+ \\
\eta^{(1)}_5 E_1 ^- & \eta^{(2)}_5E_2 ^- & \eta^{(3)}_5E_3 ^- & \eta^{(4)}_5E_1 ^+ & \eta^{(5)}_5E_2 ^+ & \eta^{(6)}_5E_3 ^+ \\
\eta^{(1)}_6E_1 ^- & \eta^{(2)}_5E_2 ^- & \eta^{(3)}_5E_3 ^- & \eta^{(4)}_5E_1 ^+ & \eta^{(5)}_5E_2 ^+ & \eta^{(6)}_5E_3 ^+ \\
\eta^{(1)}_4E_1 ^+ & \eta^{(2)}_4E_2 ^+ & \eta^{(3)}_4E_3 ^+ & \eta^{(4)}_4E_1 ^- & \eta^{(5)}_4E_2 ^- & \eta^{(6)}_4E_3 ^- \\
\eta^{(1)}_5E_1 ^+ & \eta^{(2)}_5E_2 ^+ & \eta^{(3)}_5E_3 ^+ & \eta^{(4)}_5E_1 ^- & \eta^{(5)}_5E_2 ^- & \eta^{(6)}_5E_3 ^- \\
\eta^{(1)}_6E_1 ^+ & \eta^{(2)}_6E_2 ^+ & \eta^{(3)}_6E_3 ^+ & \eta^{(4)}_6E_1 ^- & \eta^{(5)}_6E_2 ^- & \eta^{(6)}_6E_3 ^-  
\end{array} \right] \left[ \begin{array}{c} c_1 \\ c_2 \\ c_3 \\ c_4 \\ c_5 \\ c_6 \end{array} \right], 
\end{equation}
}
where $\eta^{(j)}_i, \ i,j=1,2,...,6$ is the $i$-th component of the eigenvector $\boldsymbol \eta^{(j)}$, $E_j ^{\pm} = e^{\pm p_j kh_{\mathrm d}/2}$, for $j=1,\dots,6,$ and the relation $E_{j+3} ^{\pm} = E_j^{\mp}, \ j=1,2,3$ holds.   Su et al. \cite{Su2018} showed that for the neo-Hookean, Mooney-Rivlin and Gent energy functions, the following relationships  apply
\begin{align}
\label{eigenvalue-relationship}
& \eta^{(j)}_1=-\eta^{(j+3)}_1, & &\eta^{(j)}_2=\eta^{(j+3)}_2, && \eta^{(j)}_3=-\eta^{(j+3)}_3,\nonumber\\
& \eta^{(j)}_4=\eta^{(j+3)}_4, & &\eta^{(j)}_5=-\eta^{(j+3)}_5, && \eta^{(j)}_6=\eta^{(j+3)}_6.
\end{align}


\subsection{\color{black} Bifurcation equation for wrinkles \color{black}}

The total Cauchy stress tensor used in the Stroh matrix \eqref{stroh-components} has the form
\begin{equation}
\tau_{22}=\tau_{22}^{\star}=\frac{1}{2\varepsilon_{\mathrm f}}D^2,
\end{equation}
with the incremental \textcolor{black} {interfacial conditions} at $x_2=\pm h_{\mathrm d}/2$  given by
\begin{align}\label{incremental-boundaries1}
& \dot T_{21}=-\tau_{11}^{\star}u_{2,1}+\dot{\tau}^{\star}_{21}, && \dot T_{22}=-\tau_{22}^{\star}u_{2,2}+\dot{\tau}^{\star}_{22},\nonumber\\
& \dot D_2=\dot D_2^{\star}-D^{\star}u_{2,2}, && \dot E_1=\dot E_1^{\star}+E_2^{\star}u_{2,1}.
\end{align}

It is  convenient to introduce the incremental electric potential in the silicone oil, denoted $\dot \phi^{\star}$. The nonzero components of the incremental external electric field and the corresponding incremental electric displacements   are then calculated as
\begin{equation}
\label{exterior-electricfield}
\dot E_1^{\star}= -\dfrac{\partial \dot \phi^{\star}}{\partial x_1},\qquad \dot E_2^{\star}= -\dfrac{\partial{\dot \phi^{\star}}}{\partial x_2},
\end{equation}
and
\begin{equation}
\label{exterior-displacement}
\dot D_1^{\star}=-\varepsilon_{\mathrm f} \dfrac{\partial \dot \phi^{\star}}{\partial x_1},\qquad \dot D_2^{\star}=-\varepsilon_{\mathrm f}  \dfrac{\partial\dot \phi^{\star}}{\partial x_2},
\end{equation}
respectively.

Since the applied voltage is fixed, the boundary conditions on the top and bottom faces of the tank  read
\begin{align}\label{incremental-boundaries-tank}
\dot{\phi}^{\star} (x_1, \pm H/2)=0.
\end{align}

The incremental form of Maxwell's equation outside the material simplifies to $\text{div}\dot{\mathbf D}^{\star}=0$ and the incremental electric potential  $\dot \phi^{\star}$ thus satisfies the  Laplace equation
 \begin{equation}
\label{Laplace}
\dfrac{\partial^2\dot \phi^{\star}}{\partial x_1^2} + \dfrac{\partial^2 \phi^{\star}}{\partial x_2^2} = 0.
\end{equation}

It follows that the incremental electric potential for the silicone oil in the region $H/2 \le x_2 \le h_{\mathrm d}/2$ is of the form
\begin{equation}\label{solution1}
\dot{\phi}^{\star}_+( x_1, x_2) = k^{-1}\left(C^{\star}_1e^{kx_2}+C^{\star}_2e^{-kx_2}\right)e^{\textrm{i} kx_1},
\end{equation}
where $C^{\star}_1$ and $C^{\star}_2$ are constants of integration, to be determined from the incremental interfacial and boundary conditions. 

Following Dorfmann and Ogden \cite{Dorf14a}, we write  the non-zero components of the associated incremental Maxwell stress tensor  as 
\begin{equation}
\dot{\tau}_{11}^{\star}=\dot{\tau}_{33}^{\star}=-\dot{\tau}_{22}^{\star}= D \dfrac{\partial\dot{\phi}^{\star}_{+}}{\partial x_2}, 
\qquad 
\dot{\tau}_{12}^{\star}=\dot{\tau}_{21}^{\star}= - D \dfrac{\partial \dot{\phi}^{\star}_{+}}{\partial x_1}.
\end{equation}
Substitution of Eqs.~\eqref{exterior-electric-displacement}, \eqref{incremental-fields}, \eqref{exterior-displacement} and \eqref{solution1} into the interfacial conditions Eq.~\eqref{incremental-boundaries1}$_3$ gives the connection
\begin{equation}\label{C+}
C^{\star}_1e^{kh_{\mathrm d}/2}-C^{\star}_2e^{-kh_{\mathrm d}/2}=\textrm{i} \varepsilon_{\mathrm f}^{-1}\left[DU_1\left(kh_{\mathrm d}/2\right)-\Delta\left(kh_{\mathrm d}/2\right)\right].
\end{equation}
Now substitution of Eq.~\eqref{solution1} into the boundary condition \eqref{incremental-boundaries-tank} at $x_2=H/2$ yields
\begin{equation}\label{boundary-H/2}
C^{\star}_1e^{kH/2}+C^{\star}_2e^{-kH/2}=0.
\end{equation}
Finally, by solving Eqs.~\eqref{C+} and \eqref{boundary-H/2} we obtain the two constants $C^{\star}_1$, $C^{\star}_2$ as
\begin{align}\label{C1C2}
&C^{\star}_1=\frac{\textrm{i}\left[D U_1\left(kh_{\mathrm d}/2\right)-\Delta\left(kh_{\mathrm d}/2\right)\right]e^{kh_{\mathrm d}/2}}{\varepsilon_{\mathrm f}\left(e^{kh_{\mathrm d}}+e^{kH}\right)},\nonumber\\[6pt]
&C^{\star}_2=-\frac{\textrm{i}\left[D U_1\left(kh_{\mathrm d}/2\right)-\Delta\left(kh_{\mathrm d}/2\right)\right]e^{k\left(2H+h_{\mathrm d}\right)/2}}{\varepsilon_{\mathrm f}\left(e^{kh_{\mathrm d}}+e^{kH}\right)}.
\end{align}

Using Eqs.~\eqref{exterior-electric-displacement}, \eqref{incremental-fields}, \eqref{exterior-electricfield}, \eqref{exterior-displacement} and \eqref{solution1}-\eqref{C1C2}, we  write the \color{black} remaining interfacial \color{black} conditions \eqref{incremental-boundaries1}$_{1, 2, 4}$ in an impedance form, as
\begin{equation}\label{z1}
\mathbf {S}(h_{\mathrm d}/2)= \textrm{i}\, \mathbf Z^{\star}_+ \mathbf U(h_{\mathrm d}/2),
\end{equation}
where 
\begin{equation}
\mathbf Z^{\star}_+= \varepsilon_{\mathrm f}^{-1}\left[\begin{matrix}
D^2\tanh\left[k\left(H-h_{\mathrm d}\right)/2\right]                       & -\textrm{i} D^2/2        & - D\tanh\left[k\left(H-h_{\mathrm d}\right)/2\right] \\[10pt]
\textrm{i} D^2/2     & 0                               & -\textrm{i} D \\[10pt]
- D\tanh\left[k\left(H-h_{\mathrm d}\right)/2\right]                         & \textrm{i} D               & \tanh\left[k\left(H-h_{\mathrm d}\right)/2\right]
\end{matrix}\right],
\end{equation}
is a surface impedance matrix and $\mathbf{S}(h_{\mathrm d}/2)$ and $\mathbf U(h_{\mathrm d}/2)$ are the generalised displacement and traction vectors at the face $x_2=h_{\mathrm d}/2$, respectively.

Similarly, the incremental electric potential $\dot{\phi}^{\star}$  for the silicone oil occupying the region \color{black} $-h_{\mathrm d}/2 \ge x_2 \ge -H/2$ \color{black} and satisfying the Laplace equation  \eqref{Laplace} has the form
\textcolor{black}
{\begin{equation}\label{solution2}
\dot{\phi}^{\star}_-(x_1,x_2) = k^{-1}\left(C^{\star}_3e^{kx_2}+C^{\star}_4e^{-kx_2}\right)e^{\textrm{i} kx_1},
\end{equation}}
where $C^{\star}_3$ and $C^{\star}_4$ are constants, which can be determined similarly to $C^{\star}_3$ and $C^{\star}_4$, as
\begin{align}\label{C-}
&C^{\star}_3=\frac{\textrm{i}\left[D U_1\left(-kh_{\mathrm d}/2\right)-\Delta\left(-kh_{\mathrm d}/2\right)\right]e^{k\left(2H+h_{\mathrm d}\right)/2}}{\varepsilon_{\mathrm f}\left(e^{kh_{\mathrm d}}+e^{kH}\right)},\nonumber\\[6pt] 
&C^{\star}_4=-\frac{\textrm{i}\left[D U_1\left(-kh_{\mathrm d}/2\right)-\Delta\left(-kh_{\mathrm d}/2\right)\right]e^{kh_{\mathrm d}/2}}{\varepsilon_{\mathrm f}\left(e^{kh_{\mathrm d}}+e^{kH}\right)}.
\end{align}

The interfacial conditions on the plate's face at $x_2=-h_{\mathrm d}/2$ can then be rewritten as
\begin{equation}\label{z2}
\mathbf {S}(-kh_{\mathrm d}/2) =\textrm{i}  {\mathbf Z}^{\star}_-\mathbf U(-kh_{\mathrm d}/2),
\end{equation}
where
\begin{equation}
\mathbf Z^{\star}_-= \varepsilon_{\mathrm f}^{-1}\left[\begin{matrix}
-D^2\tanh\left[k\left(H-h_{\mathrm d}\right)/2\right]                       & -\textrm{i} D^2/2        & D\tanh\left[k\left(H-h_{\mathrm d}\right)/2\right] \\[10pt]
\textrm{i} D^2/2     & 0                               & -\textrm{i} D \\[10pt]
D\tanh\left[k\left(H-h_{\mathrm d}\right)/2\right]                         & \textrm{i} D               & -\tanh\left[k\left(H-h_{\mathrm d}\right)/2\right]
\end{matrix}\right],
\end{equation}
is a surface impedance matrix and $\mathbf{S}(-kh_{\mathrm d}/2)$ and $\mathbf U(-kh_{\mathrm d}/2)$  are the generalised displacement and traction vectors on the face $x_2=-h_{\mathrm d}/2$, respectively. 

Combining Eqs.~\eqref{generalised-displacement}, \eqref{z1} and \eqref{z2} gives
{\footnotesize
\begin{align}\label{full}
& \left[ \begin{array}{c} \mathbf{S}(kh/2) \\
\mathbf S(-kh/2) \end{array} \right]   =
\textrm{i} \left[\begin{matrix}
\mathbf Z^{\star}_+ & \mathbf 0\\
\mathbf 0 &  {\mathbf Z}^{\star}_-
\end{matrix}\right]
\left[ \begin{array}{c} \mathbf{U}(kh/2) \\
\mathbf U(-kh/2) \end{array} \right] \nonumber\\[6pt]
& \qquad = \textrm{i}
\left[\begin{matrix}
\mathbf Z^{\star}_+ & \mathbf 0\\
\mathbf 0 &  {\mathbf Z}^{\star}_-
\end{matrix}\right]
\left[ \begin{array}{cccccc} 
\eta^{(1)}_1 E_1 ^- & \eta^{(2)}_1 E_2 ^- & \eta^{(3)}_1 E_3 ^- & \eta^{(4)}_1 E_1 ^+ & \eta^{(5)}_1E_2 ^+& \eta^{(6)}_1 E_3 ^+\\
\eta^{(1)}_2 E_1 ^- & \eta^{(2)}_2E_2 ^-& \eta^{(3)}_2E_3 ^- & \eta^{(4)}_2E_1 ^+ & \eta^{(5)}_2E_2 ^+ & \eta^{(6)}_2E_3 ^+ \\
\eta^{(1)}_3E_1 ^- & \eta^{(2)}_3E_2 ^-& \eta^{(3)}_3E_3 ^-& \eta^{(4)}_3E_1 ^+ & \eta^{(5)}_3E_2 ^+ & \eta^{(6)}_3E_3 ^+ \\
\eta^{(1)}_1E_1 ^+ & \eta^{(2)}_1E_2 ^+ & \eta^{(3)}_1E_3 ^+ & \eta^{(4)}_1E_1 ^- & \eta^{(5)}_1E_2 ^- & \eta^{(6)}_1E_3 ^- \\
\eta^{(1)}_2E_1 ^+ & \eta^{(2)}_2E_2 ^+ & \eta^{(3)}_2E_3 ^+ & \eta^{(4)}_2E_1 ^- & \eta^{(5)}_2E_2 ^- & \eta^{(6)}_2E_3 ^- \\
\eta^{(1)}_3E_1 ^+ & \eta^{(2)}_3E_2 ^+ & \eta^{(3)}_3E_3 ^+ & \eta^{(4)}_3E_1 ^- & \eta^{(5)}_3E_2 ^- & \eta^{(6)}_3E_3 ^-  \end{array} \right] \left[ \begin{array}{c} c_1 \\ c_2 \\ c_3 \\ c_4 \\ c_5 \\ c_6 \end{array} \right].
\end{align}
}

The decoupled bifurcation equations of the plate are then obtained from simple algebraic manipulations  \cite{Nayfeh95} of Eqs.~\eqref{generalised-stress} and \eqref{full}. 
Considering antisymmetric modes only, we find
\begin{equation}
\label{asymmetric2}
 \left |\begin{array}{cccc}
 P_{11} & P_{12} & P_{13} \\
 P_{21} & P_{22} & P_{23} \\
 P_{31} & P_{32} & P_{33} 
 \end{array} \right |
 =0,
\end{equation}
where
{\small
\begin{align}
&P_{1j}=\eta ^{(j)}_4-\frac{D}{2\varepsilon_{\mathrm f}}\left\{D\eta ^{(j)}_2+2\textrm{i}\left(-D\eta ^{(j)}_1+\eta ^{(j)}_3\right)\tanh\left[k(H-h_{\mathrm d})/2\right]\tanh(p_jkh_{\mathrm d}/2)\right\},\nonumber\\[10pt]
&P_{2j}=\left[\eta ^{(j)}_5+\frac{D}{2\varepsilon_{\mathrm f}}\left(D\eta ^{(j)}_1-2\eta ^{(j)}_3\right)\right]\tanh(p_jkh_{\mathrm d}/2),\nonumber\\[10pt]
&P_{3j}=\eta ^{(j)}_6+\frac{1}{\varepsilon_{\mathrm f}}\left\{D\eta ^{(j)}_2+\textrm{i}\left(-D\eta ^{(j)}_1+\eta ^{(j)}_3\right)\tanh\left[k(H-h_{\mathrm d})/2\right]\tanh(p_jkh_{\mathrm d}/2)\right\}.
\end{align}
}

\color{black}
Eq. \eqref{asymmetric2} is the \textit{bifurcation criterion for the antisymmetric wrinkling mode of instability}, which always occurs first  \cite{Dorf14a, Dorfmann2019, Su2018}.
\color{black}


\section{Numerical results for neo-Hookean dielectric plates}
\label{section4}


Here we specialise the theory presented in Section \ref{section3} to equi-biaxial deformations \color{black} ($\lambda_1 = \lambda_3 = \lambda$ and $s_1=s_3=s$) \color{black}  of a \emph{neo-Hookean electro-elastic} plate. Full details  are given in the Appendix.

\subsection{Static response}


The nonlinear response of an immersed neo-Hookean electro-elastic  plate subject to a biaxial deformation is given  in terms of the non-dimensional in-plane nominal stresses $\bar s_1=s_1/\mu$, $\bar s_2=s_2/\mu$ and of the electric potential in the Appendix, see Eqs. \eqref{VE2} and \eqref{nonlinear-response-biaxial-2}. 
Specialising to equi-biaxial deformation, i.e. $\bar s_1=\bar s_2=\bar s$ and $\lambda_1=\lambda_3=\lambda$, we obtain
\begin{equation}
\label{loading2a}
\bar  V=\left[\bar  H\left(1-\bar  \varepsilon\right)+\bar  \varepsilon\lambda^2\right]
\sqrt{\frac{\left(\bar  s-\lambda\right)\lambda^5+1}{\left(\bar \varepsilon-1\right)\lambda^8}},
\end{equation}
where \color{black} $\bar  V=(V/H)\sqrt{\varepsilon_{\mathrm d}/\mu}$ is a non-dimensional measure of the applied voltage, $\mu$ is the shear modulus of the solid in the absence of voltage, \color{black} $\bar  H=H_{\mathrm d}/H$ and $\bar  \varepsilon=\varepsilon_{\mathrm d}/\varepsilon_{\mathrm f}$  are the ratios of the initial plate thickness to the distance of the electrodes and  of the plate permittivity  to that of the silicone oil, respectively. 
The voltage-induced stretch of the plate $\lambda_{\mathrm V}$ is then obtained from  Eq. \eqref{loading2a} for $\bar  s=0$.

\color{black} It is well established that a pull-in instability of a neo-Hookean dielectric plate may be triggered when the nonlinear $\bar V-\lambda$ response curve reaches a peak point \cite{ZhSu07, Su2018b}. \color{black}
\color{black}
From Eq. \eqref{loading2a}, $\bar s$ corresponding to the peak point $\text d \bar V/\text d \lambda=0$ is obtained as
\begin{equation}\label{ssss}
\bar s_{\text{cr}}=\frac{2\left[\bar H\left(\bar\varepsilon-1\right)\left(\lambda^6-4\right)+\bar\varepsilon\lambda^2\left(\lambda^6+2\right)\right]}{\lambda^5\left[3\bar H\left(\bar\varepsilon-1\right)+\bar\varepsilon\lambda^2\right]}.
\end{equation}
Inserting Eq. \eqref{ssss} into Eq. \eqref{loading2a} then yields the \textit{Hessian criterion} as\color{black}
\begin{equation}
\label{Hessian2}
\bar  V=\sqrt\frac{\left[\bar  H\left(1-\bar  \varepsilon\right)+\bar  \varepsilon\lambda^2\right]^3\left(\lambda^6+5\right)}{\left(\bar \varepsilon-1\right)
\left[3\bar  H\left(\bar  \varepsilon-1\right)+\bar  \varepsilon\lambda^2\right]
\lambda^8}.
\end{equation}

Eqs.~\eqref{loading2a} and \eqref{Hessian2} show that the  response of the electro-elastic plate depends on the values of  $\bar  H$ and $\bar \varepsilon$. 
In particular, \color{black} from Eq. \eqref{loading2a} we can see that when \color{black} $\bar  \varepsilon=1$, the in-plane nominal stress $\bar s$ is not influenced by $\bar V$ and the  plate deformation is independent of the electric field.  
When $\bar  \varepsilon=0$,  $H_{\mathrm d}=H$, \color{black} Eq.~\eqref{loading2a}  recovers the   equation governing the static deformation of DE plates with no exterior electric field, and  \eqref{Hessian2} recovers its associated Hessian criterion \cite{ZhSu07}. \color{black}

\begin{figure}[b!]
\centering
\includegraphics[width=0.48\textwidth]{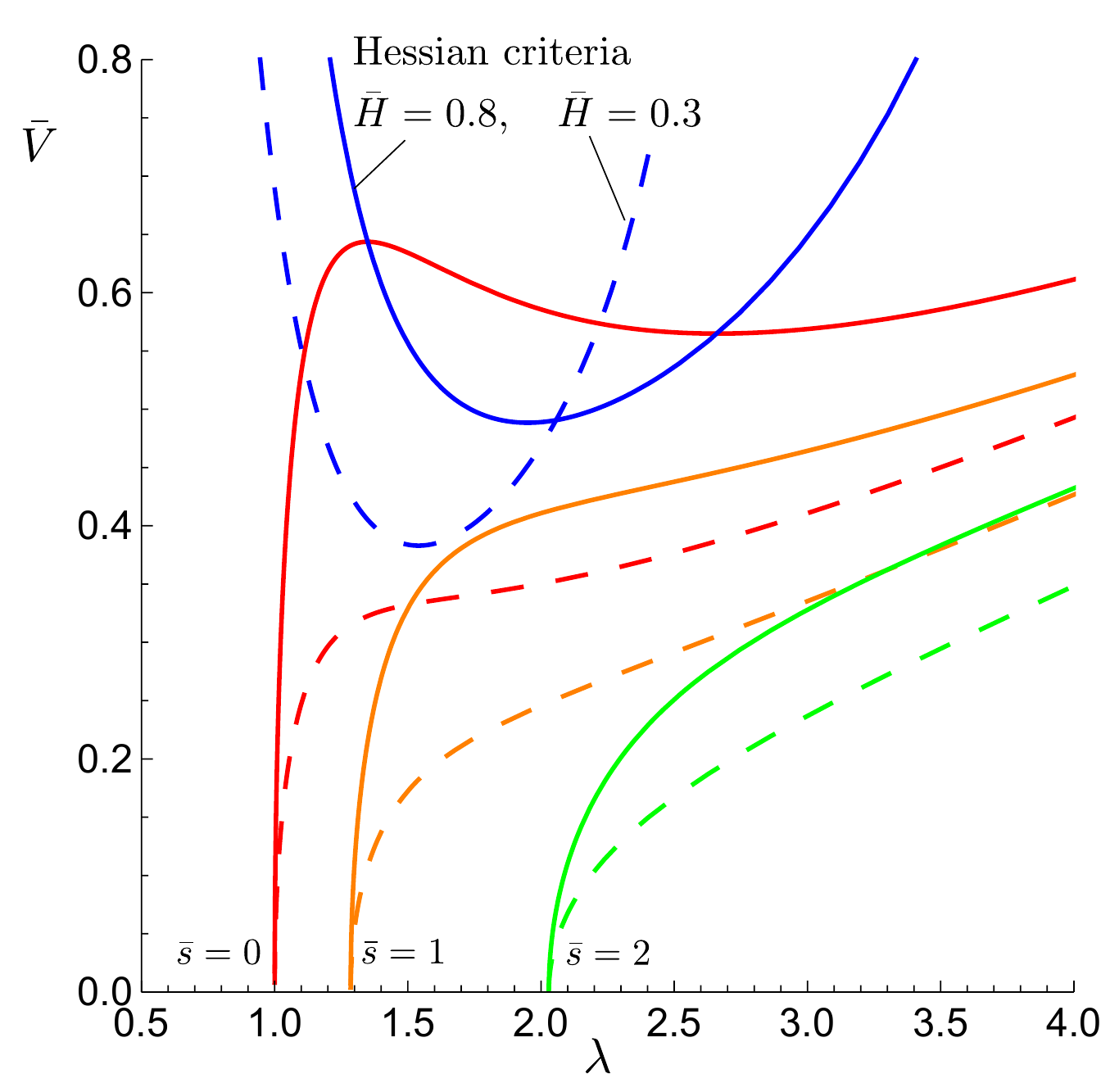}
\includegraphics[width=0.48\textwidth]{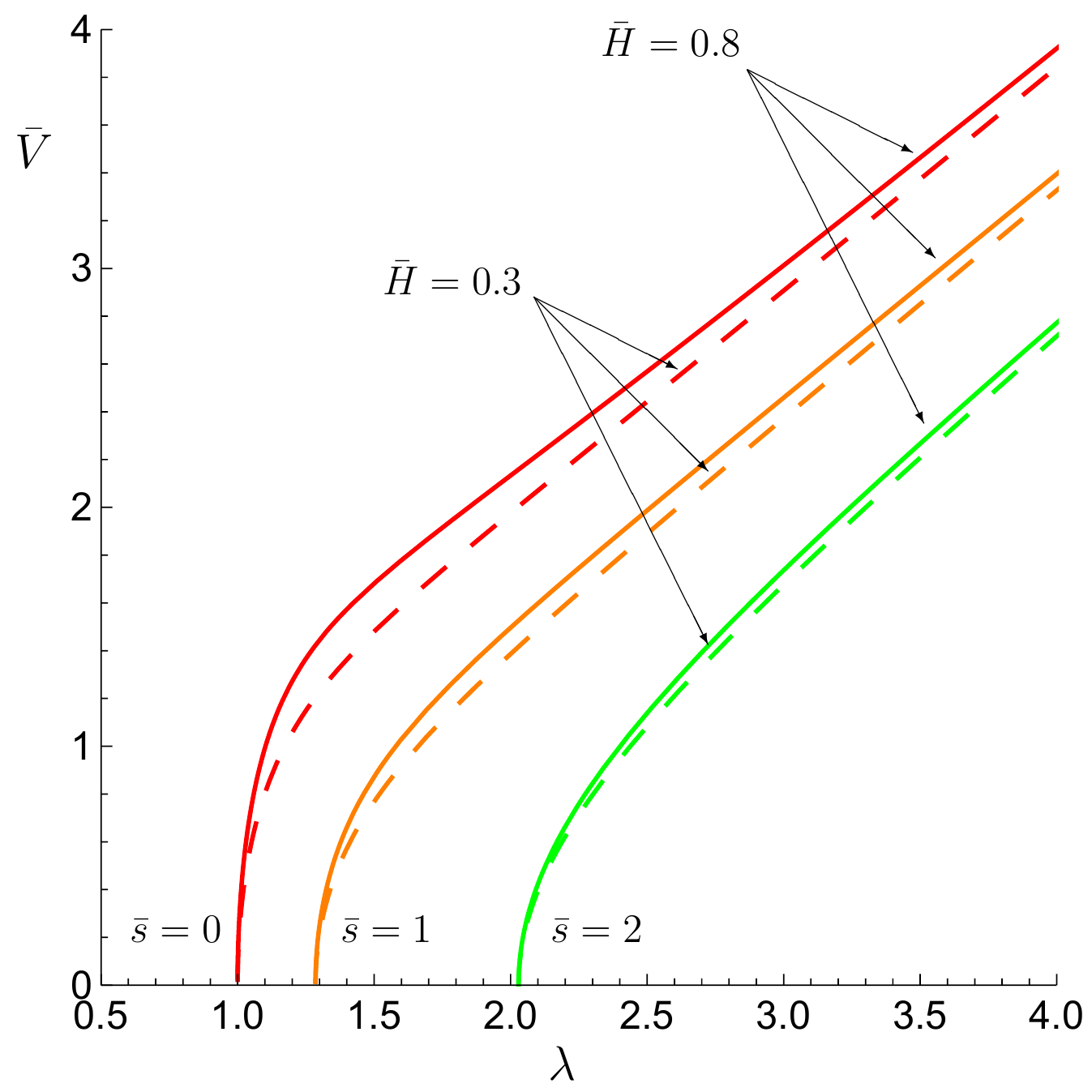}
\caption{
{\footnotesize
The continuous and dashed curves represent the  nonlinear  responses of a \color{black} neo-Hookean \color{black}  electro-elastic plate immersed in silicone oil, for  $\bar H=0.8$ and $\bar H=0.3$, respectively. 
    The non-dimensionalised values of the equi-biaxial in-plane stress are  $\bar  s=0,1,2$ and  correspond to the red, orange and green curves, respectively. 
The image on the left shows the Hessian  criterion \color{black} and the loading curves for a permittivity ratio  $\bar  \varepsilon=0.1$. 
    The image on the right depicts the  $\bar V-\lambda$ curves for $\bar  \varepsilon=0.6$. 
    The Hessian criteria curves do not cross any of the loading curves and are therefore not included in the latter.}
}
\label{figure5}
\end{figure}

The left image in Figure \ref{figure5} displays the behaviour of an electro-elastic neo-Hookean plate immersed in silicone oil with relative permittivity $\bar \varepsilon=0.1$, for relative thickness values $\bar H=0.3$ (dashed curves) and $\bar H=0.8$ (solid  curves).  
The non-dimensional in-plane pre-stresses $\bar s=0, 1,2$ are considered, shown by red, orange and green curves, respectively. 
The curves corresponding to the Hessian criterion when $\bar H=0.3$ (dashed curve) and $\bar H=0.8$ (solid  curve) are also displayed.  
As an electric field is applied,  the stretch increases until a homogeneous pull-in instability 
occurs when the loading curve crosses the Hessian criterion. \color{black} This occurs, for example, for $\bar s = 0$ and $\bar H=0.8$, but not for $\bar s = 0$ and $\bar H=0.3$. \color{black}
The image on the right shows the results corresponding to $\bar \varepsilon=0.6$\color{black}. I\color{black}n that case, the curves corresponding to the Hessian \color{black} criteria \color{black} do not intersect  any of the three loading curves and for the sake of clarity are therefore not shown. 
The figures show that the pull-in instability can be suppressed by increasing the value of $\bar\varepsilon$ and\color{black}/or \color{black} reducing  the value of $\bar H$. 
They also show that increasing values of $\bar\varepsilon$ and/or $\bar H$ stiffen the response, i.e. a larger potential is needed to achieve the same deformation. 

\begin{figure}[t!]
\centering
\includegraphics[width=1.0\textwidth]{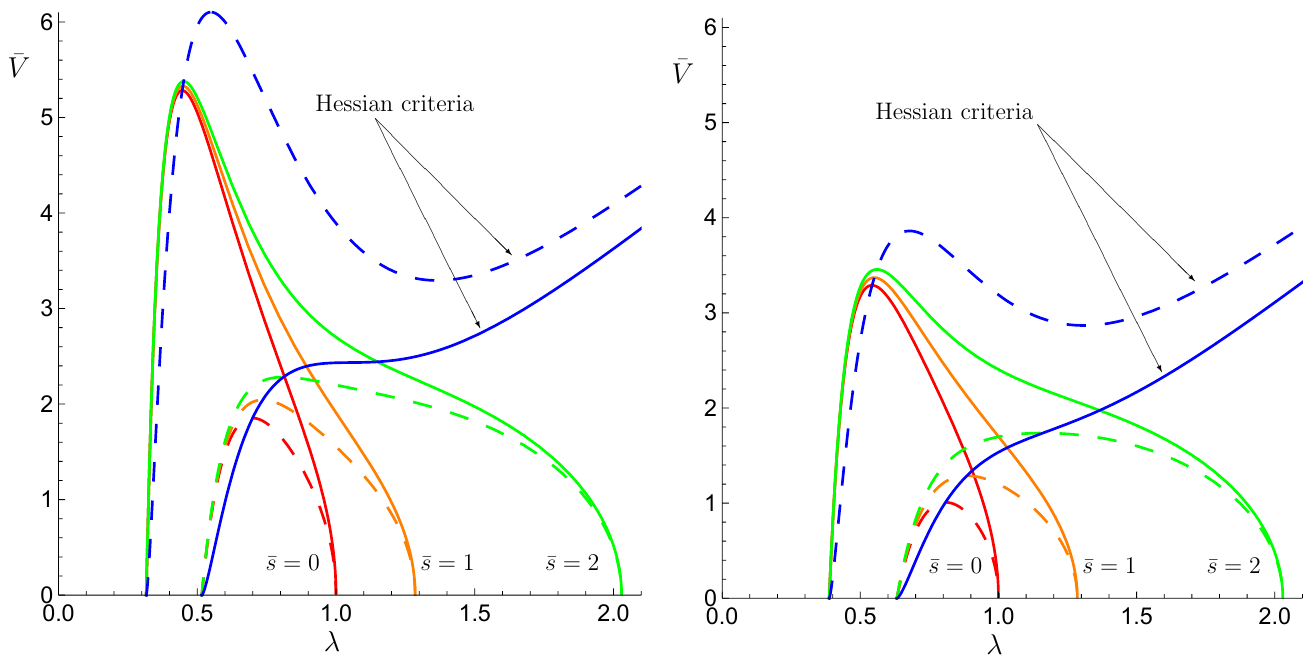}
\caption{
{\footnotesize
The Hessian \color{black} criterion \color{black} and the nonlinear response curves of an immersed neo-Hookean electro-elastic plate  with  $\bar H=0.8$ and $\bar H=0.3$  are depicted by continuous and dashed curves, respectively. The equi-biaxial in-plane stresses, in non-dimensionalised form, are  $\bar  s=0,1,2$ and  correspond, respectively, to the red, orange and green curves. The image on the left represents the behaviour for  $\bar  \varepsilon=1.5$, the one on the right for  $\bar  \varepsilon=2$.}
}
\label{figure6}
\end{figure}

The dependence of the in-plane stretch $\lambda$ on the non-dimensional form of the electric potential  $\bar V$ for values of  $\bar \varepsilon=1.5, 2$   is shown in Figure \ref{figure6}. 
In contrast  to the behaviour depicted in Figure \ref{figure5}, the application of  an electric field now induces a \textit{lateral contraction of the plate}, as $\lambda$ decreases with increasing $\bar V$. 
The initial response is monotonic until a critical point is reached when $\mathrm d \bar V/\mathrm d \lambda =0$. At this point the Hessian criterion indicates  a  \color{black} ``\textit{pull-out''  instability}, \color{black} with \color{black} a uniform increase in thickness  accompanied by large lateral contraction. \color{black} The system becomes more stable for lower values of $\bar H$ and $\bar \varepsilon$.

\subsection{\color{black}Wrinkling \color{black}  analysis}
Specialising Eq.~\eqref{bifurcation2} to equi-biaxial deformation, and  using Eq. \eqref{VE2}, gives \color{black} the \color{black}wrinkling \color{black}  criterion for the immersed plate as 
\color{black}
\begin{align}
\label{equibiaxial-buckling2}
\bar  V=&\left[\bar  H\left(1-\bar  \varepsilon\right)+\bar  \varepsilon\lambda^2\right]
\nonumber \\[4pt]
&\quad  \sqrt{\bar \varepsilon\tanh\left[\pi \left(\bar H \lambda^{2}-1\right)H_{\mathrm d}/\left(\lambda^2\mathcal L\right)\right]+\coth\left[\pi H_{\mathrm d}/\left(\lambda^2\mathcal L\right)\right]}  
\nonumber\\[4pt]
&\quad \quad \sqrt{\frac{\left(\lambda^6+1\right)^2\tanh\left[\pi H_{\mathrm d}/\left(\lambda^2\mathcal L\right)\right]-4\lambda^3\tanh\left(\pi  \lambda H_{\mathrm d}/\mathcal L\right)}{\lambda^8\left(\bar \varepsilon-1\right)^2\left(\lambda^6-1\right)}}.
\end{align}  
\color{black}

The \color{black}wrinkling \color{black}  criteria \color{black} for thin- and thick-plates are obtained by evaluating \eqref{equibiaxial-buckling2} when  $H_{\mathrm d}/\mathcal L\rightarrow 0$ \color{black} and  $H_{\mathrm d}/\mathcal L \rightarrow \infty$\color{black}, respectively. 
For thin plates we have 
\begin{equation}
\label{equibiaxial-thin-plate2}
\bar  V=\left[\bar  H\left(1-\bar  \varepsilon\right)+\bar  \varepsilon\lambda^2\right]\sqrt{\frac{\lambda^6-1}{\left(\bar \varepsilon-1\right)^2\lambda^8}},
\end{equation}
while for thick plates we have
\begin{equation}
\label{equibiaxial-thick-plate2}
\bar  V=\left[\bar  H\left(1-\bar  \varepsilon\right)+\bar  \varepsilon\lambda^2\right]\sqrt{\frac{\left(\bar \varepsilon+1\right)\left(\lambda^9+\lambda^6+3\lambda^3-1\right)}{\lambda^8\left(\bar \varepsilon-1\right)^2\left(\lambda^{3}+1\right)}}.
\end{equation}

Note that \color{black} unlike the results in \cite{Su2018} (no exterior electrical field), \color{black} the loading curve $\bar  V-\lambda$ with no in-plane pre-stresses is not equivalent to the \color{black}wrinkling \color{black} limit for plates with vanishing thickness.
\begin{figure}[b!]
\centering
\includegraphics[width=1.0\textwidth]{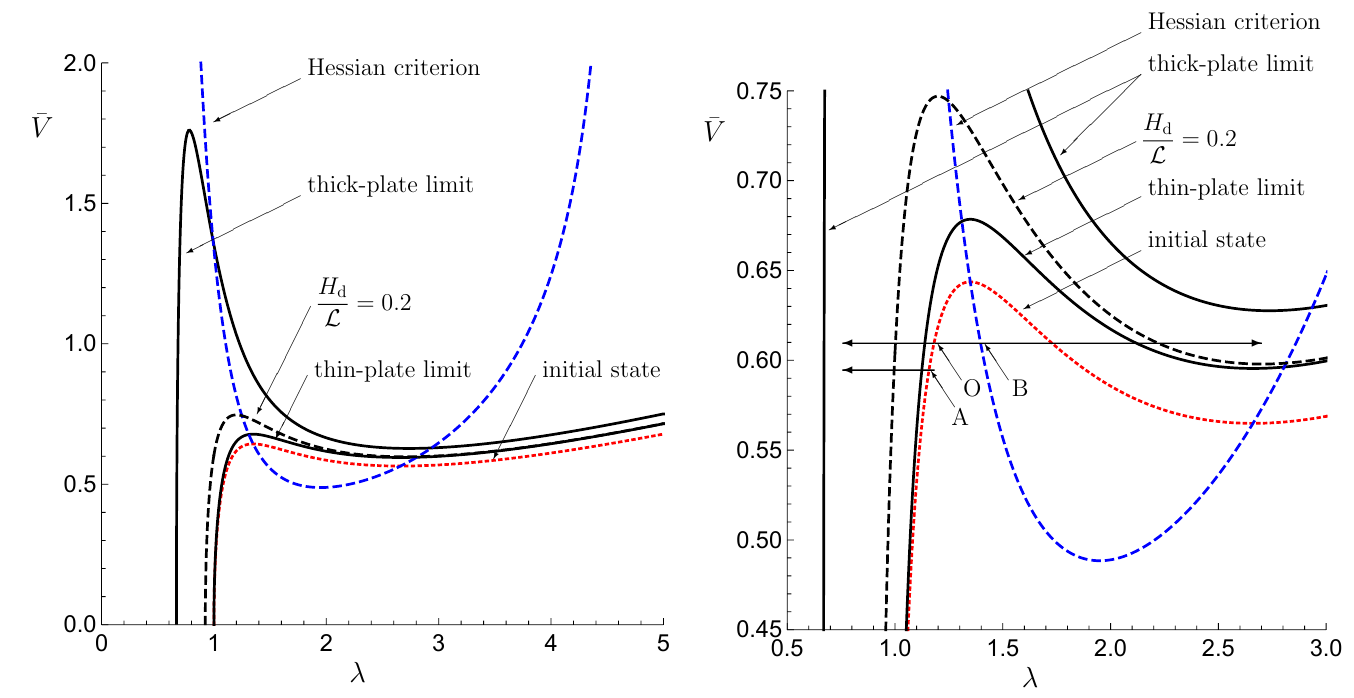}
\caption{
{\footnotesize
The \color{black}wrinkling \color{black}  behaviour of an immersed neo-Hookean electro-elastic plate with $\bar H=0.8$ and $\bar \varepsilon=0.1$. 
The solid  curves correspond to the thin- and thick-plate limits, the   dashed black curve represents the behaviour when   \color{black} $H_{\mathrm d}/\mathcal L=0.2$\color{black}.  The initial state is obtained by the application of an electric potential with $\bar s=0$ and is depicted by a red dotted curve. It shows that an exterior electric field has a stabilising effect.   
The Hessian criterion indicates the onset of a pull-in instability and is depicted by the blue dashed curve.}
}
\label{figure7}
\end{figure}

\begin{figure}[b!]
\centering
\includegraphics[width=1.0\textwidth]{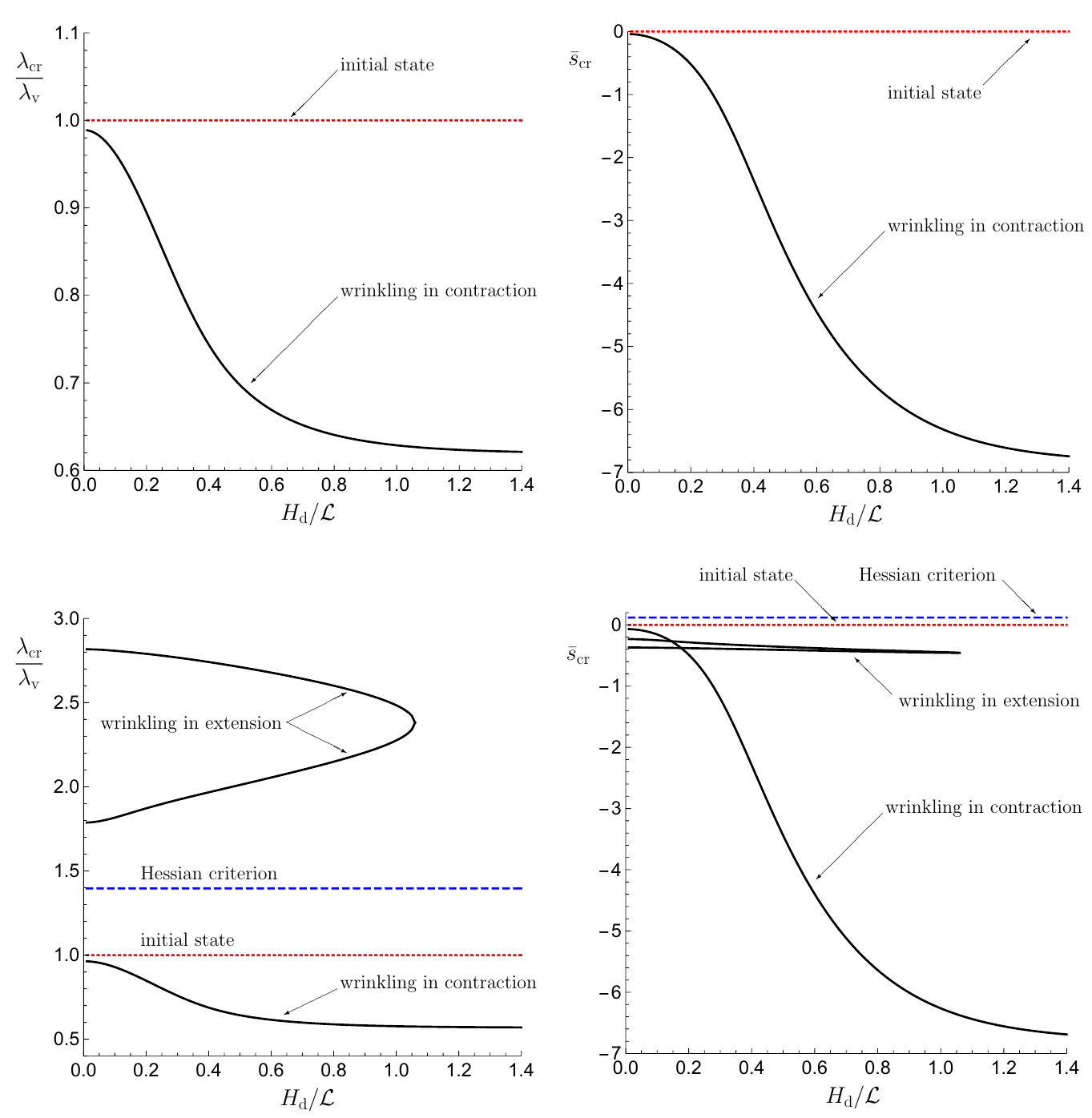}
\caption{
{\footnotesize 
The upper and lower images show the ratio $\lambda_{\mathrm{cr}}/ \lambda_{\mathrm V}$ and the  non-dimensional in-plane critical stress $\bar s_{\mathrm{cr}}$ as a function of the thickness to \color{black} wavelength \color{black}  ratio \color{black} $H_{\mathrm d}/\mathcal L$ \color{black}  for the electric potentials $\bar V=0.5$ and $\bar V=0.61$, respectively
\color{black}(plate immersed in silicone oil)\color{black}.  The solid curves  represent the \color{black}wrinkling \color{black}  behaviour in contraction and extension. The red dotted  and blued dashed curves show the initial state and the Hessian criterion, respectively. 
}
}
\label{figure8}
\end{figure}

Figures \ref{figure7}  and \ref{figure8} summarise the response of an immersed neo-Hookean electro-elastic plate  with relative permittivity $\bar \varepsilon=0.1$ and thickness ratio $\bar  H=0.8$. 
Figure \ref{figure7}, in particular, depicts the Hessian criterion \eqref{Hessian2}, the wrinkling criterion \eqref{equibiaxial-buckling2} for \color{black} $H_{\mathrm d}/\mathcal L=0.2$ \color{black} and the  criteria  \eqref{equibiaxial-thin-plate2}, \eqref{equibiaxial-thick-plate2} for  thin and thick plates, respectively. 
The initial state, depicted by a red dashed curve is obtained from \eqref{loading2a} with $\bar s=0$. 
It shows that the electric field generated by $\bar V$ induces a reduction in  thickness and by incompressibility, an equi-biaxial stretch $\lambda$. 
Consider, for example,  the initial state $\mathrm O$ of a thin plate with \color{black} $H_{\mathrm d}/\mathcal L\rightarrow 0$ \color{black}  defined by point $\mathrm O$ in the right image in Figure \ref{figure7}. 
It shows that an in-plane biaxial compression $\bar s<0$ is required to induce the wrinkled state. 
This differs from the behaviour of \color{black} DE plates shown in Figure \ref{figure1}a by Su et al. \cite{Su2018} \color{black}, where a thin plate buckles immediately upon the application of a small compressive mechanical load. Alternatively,  an in-plane tensile stress deforms the plate  from the initial state $\mathrm O$ to point $\mathrm B$ where the Hessian criterion is met, resulting in uncontrolled uniform thinning. Of interest is the initial state defined by point $\mathrm A$ and corresponding to an electric potential \color{black} $\bar V=0.59$. \color{black} It shows that \color{black}wrinkling \color{black}  occurs for in-plane contraction with the magnitude a function of the ratio \color{black} $H_{\mathrm d}/\mathcal L$\color{black}. Maintaining \color{black} $\bar V=0.59$ \color{black} constant, an in-plane extension of sufficient magnitude will satisfy the Hessian criterion but  \color{black}wrinkling \color{black}  in tension will not occur.

Figure \ref{figure8} shows the ratio $\lambda_{\mathrm{cr}}/ \lambda_{\mathrm V}$ and the  non-dimensional in-plane critical stress $\bar s_{\mathrm{cr}}$ as a function of the thickness to \color{black} wavelength \color{black}  ratio \color{black} $H_{\mathrm d}/\mathcal L$ \color{black}  for the electric potentials $\bar V=0.5$ and $\bar V=0.61$ using  solid curves  to indicate wrinkling \color{black} (here $\lambda_{\mathrm{cr}}$ is the critical stretch of wrinkling\color{black}).  In particular, for $\bar V=0.5$ wrinkling occurs in compression only with in-plane contraction and critical stress increasing with \color{black} $H_{\mathrm d}/\mathcal L$\color{black}. 
For an electric potential $\bar V=0.61$   the behaviour in  compression is similar; however, the behaviour in extension is different, with  equi-biaxial deformation of sufficient magnitude inducing uniform thinning when the Hessian criterion is met.   
For increasing values of $\lambda$, \color{black}wrinkling \color{black}  in extension occurs.

\begin{figure}[b!]
\centering
\includegraphics[width=1.0\textwidth]{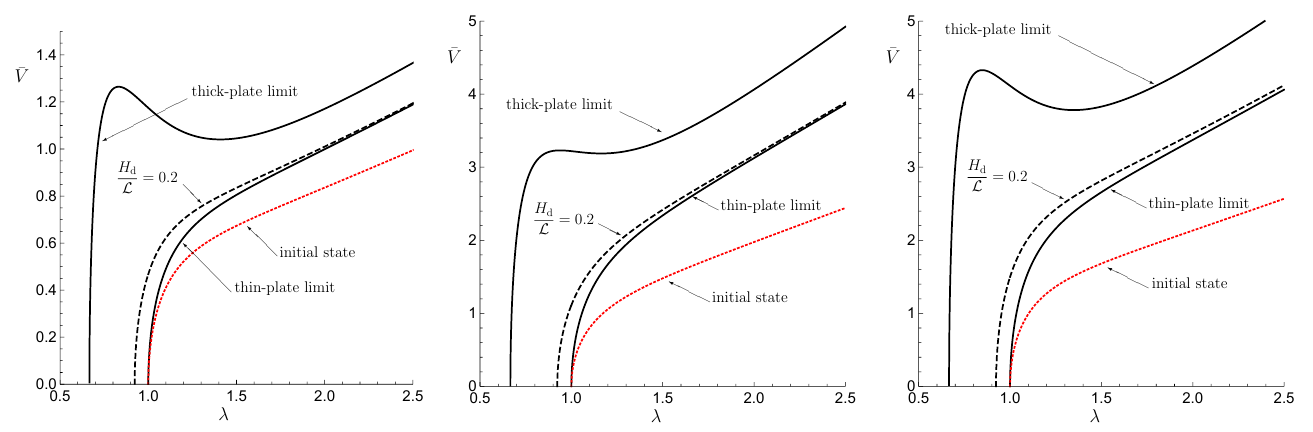}
\caption{
{\footnotesize
Dependence of the non-dimensional electric potential $\bar V$ on the in-plane stretch $\lambda$ for  values of  $\bar \varepsilon=0.3,0.6$ and  $\bar H=0.3,0.8$.
Specifically, the image on the left shows the \color{black}wrinkling \color{black}  behaviour and the initial state for $\bar \varepsilon=0.3, \bar H=0.3$, the centre image corresponds to $\bar \varepsilon=0.6, \bar H =0.3$ and the results on the right to  $\bar \varepsilon=0.6, \bar H=0.8$. }
}
\label{figure9}
\end{figure}

The initial state, the \color{black}wrinkling \color{black}  behaviour for a plate with \color{black} $H_{\mathrm d}/\mathcal L=0.2$, \color{black}  and  the thin- and thick-plate limits are shown in Figure \ref{figure9} for $\bar\varepsilon=0.3,0.6$ and for $\bar H=0.3, 0.8$. 
The initial state for a potential difference $\bar V$ is defined by stable biaxial stretch $\lambda$ with magnitude influenced by $\bar\varepsilon$ and $\bar H$. 
Take, for example,  the initial state defined by $\lambda=2$. The two images on the left show that the required potential difference  increases with $\bar\varepsilon$. 
On the other hand, the two images on the right show that a change in  $\bar H$ does not  significantly influence the value of $\bar V$ to obtain the initial deformation $\lambda=2$. 
A subsequently applied in-plane biaxial stress  $\bar s < 0$, with  $\bar V$ constant,  induces \color{black}wrinkling \color{black}  with the amount of contraction increasing with \color{black} $H_{\mathrm d}/\mathcal L$\color{black}. Alternatively, to keep  $\lambda=2$ constant when the potential $\bar V$  is increased requires an in-plane stress $\bar s<0$.  The graphs show that  the increase in $\bar V$  \color{black} inducing \color{black} \color{black}wrinkling\color{black}, for $\lambda=2$  constant,  depends on $\bar \varepsilon$ and that the influence of $\bar H$ is minor. 

\begin{figure}[t!]
\centering
\includegraphics[width=0.5\textwidth]{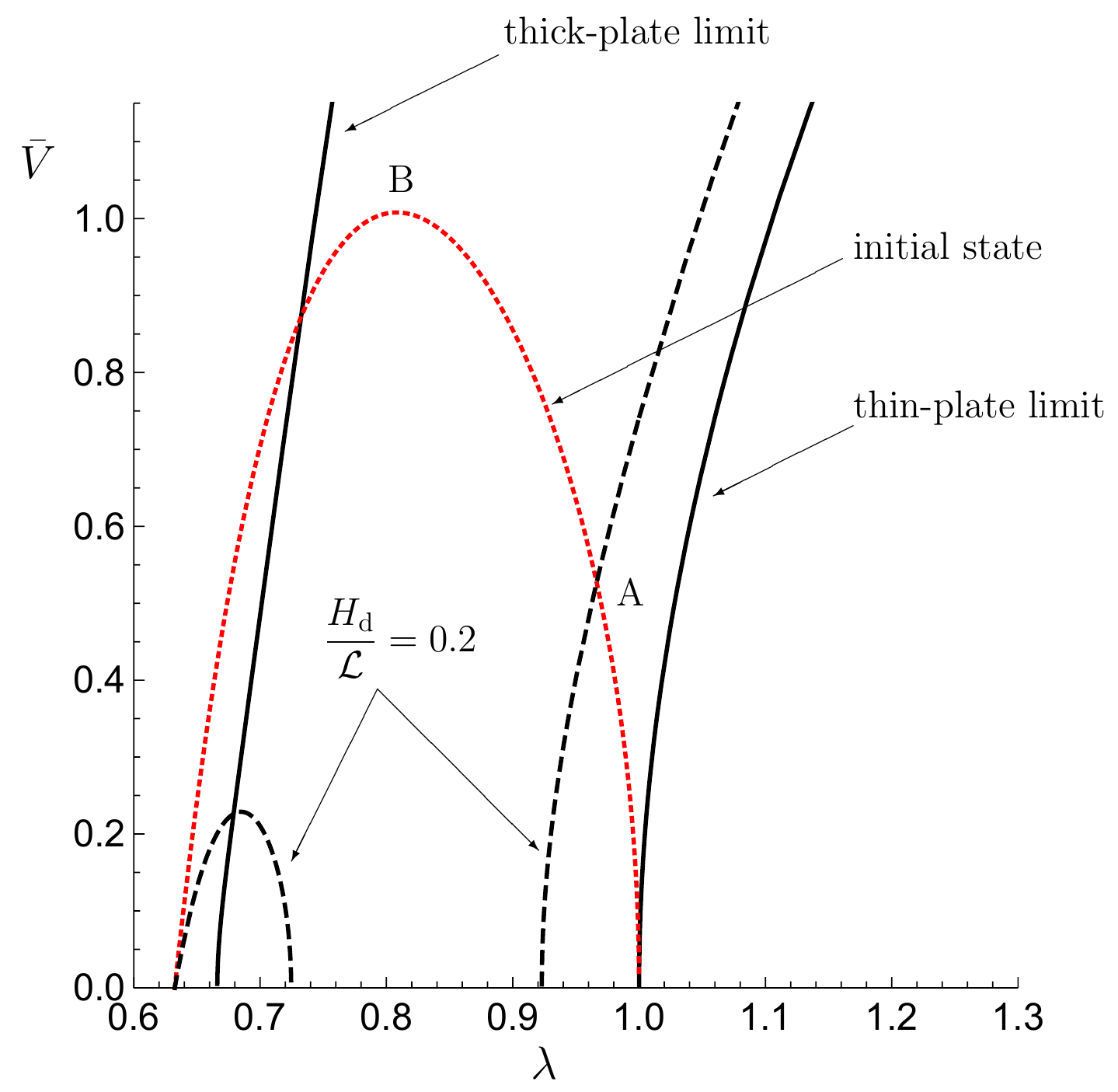}
\caption{
{\footnotesize
The $\bar V-\lambda$ loading curves and the initial state of an immersed neo-Hopokean electro-elastic plate for  permittivity and  thickness ratios $\bar \varepsilon=2$ and $\bar  H=0.8$, respectively.  The solid curves depict the behaviour for the  thin- and thick-plate limits, the dashed curve corresponds to \color{black} $H_{\mathrm d}/\mathcal L=0.2$ \color{black}  and the dotted red curve represents the initial state when  $\bar s=0$.}
}
\label{figure10}
\end{figure}


Figures \ref{figure10}  illustrates the $\bar V-\lambda$ loading curves of an immersed electro-elastic plate for a relative permittivity $\bar \varepsilon=2$ and for a ratio of plate thickness to electrode distance  $\bar H= 0.8$. 
In addition to the \color{black}wrinkling \color{black}  behaviour for  thin- and thick-plate limits and for \color{black} $H_{\mathrm d}/\mathcal L=0.2$\color{black},  Figure \ref{figure10} also  depicts the initial state obtained by a potential difference at the electrodes with $\bar s=0$. As already shown in Figure \ref{figure6}, the potential difference at the electrodes, with $\bar s=0$,  induces  an increase in plate thickness and by incompressibility a biaxial contraction $\lambda<1$. The initial state of a plate with \color{black} $H_{\mathrm d}/\mathcal L=0.2$\color{black}, for example, buckles at point $\mathrm A$ and can be stabilised by the application of an in-plane biaxial tension $\bar s>0$. For plates with  \color{black} $H_{\mathrm d}/\mathcal L>0.2$ \color{black}  and with the electric potential constant and equal to the one indicated by point $\mathrm A$, an in-plane compression $\bar s<0$ is necessary to induce a wrinkled configuration. The electric potential $\bar V$ corresponding to the initial state $\mathrm B$ induces a \color{black} pull-out \color{black} instability in the plate, see Figure \ref{figure6}.

\begin{figure}[b!]
\centering
\includegraphics[width=1.0\textwidth]{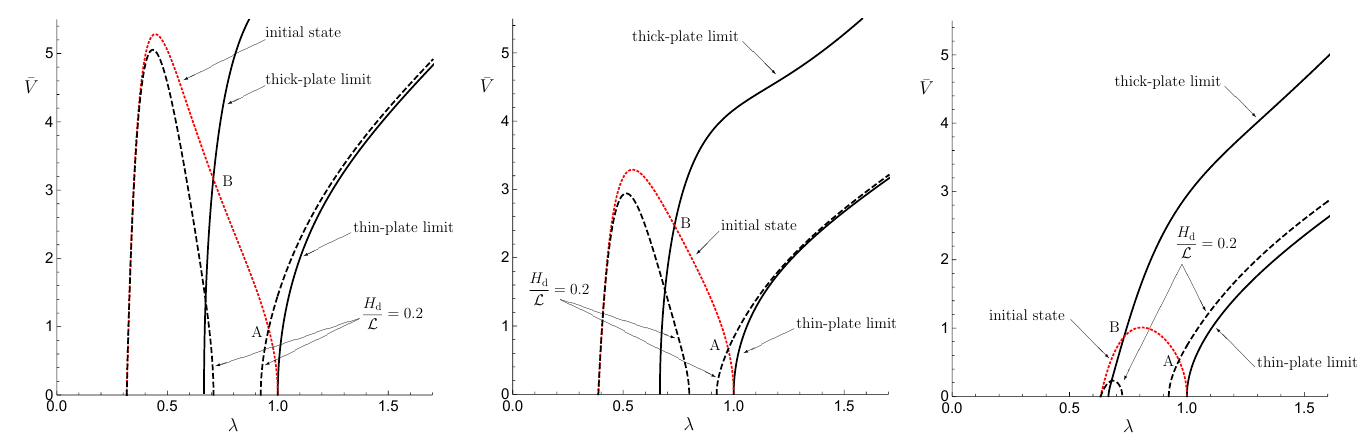}
\caption{
{\footnotesize
The image on the left shows the initial state and the dependence of the non-dimensional electric potential $\bar V$ on the in-plane stretch $\lambda$   for values of $\bar \varepsilon=1.5$ and $\bar H=0.3$. The image in the centre is obtained when $\bar \varepsilon=2$ and $\bar H=0.3$ and the results on the right correspond to $\bar \varepsilon=2$ and $\bar H=0.8$.}
}
\label{figure12}
\end{figure}

The \color{black}wrinkling \color{black}  behaviour for the thin- and thick-plate limits and for \color{black} $H_{\mathrm d}/\mathcal L=0.2$ \color{black}   as well as the initial state are shown in Figure \ref{figure12} for $\bar\varepsilon=1.5,2$ and $\bar H=0.3,0.8$. The main difference with the results in Figure \ref{figure9} is the use of larger values of $\bar \varepsilon$.  The results indicate that  increasing the potential difference at the electrodes, with in-plane stress $\bar s=0$, induces wrinkles at point $\mathrm A$ for a plate  with   \color{black} $H_{\mathrm d}/\mathcal L=0.2$ \color{black}  and at point  $\mathrm B$ for the thick-plate limit.  
In particular, consider the initial state defined by a biaxial stretch $\lambda=0.8$. 
The three images  show a reduction in the electric potential with increased values of  $\bar\varepsilon$ and $\bar H$.  An in-plane contraction or extension, superposed on the initial state $\lambda=0.8$, with $\bar V$ constant,  induces a wrinkled configuration  depending on the specific value of $H_{\mathrm d}/L$. Figure \ref{figure12} also shows that an in-plane tension $\bar s>0$ must be superposed to keep  the stretch $\lambda=0.8$ constant for increasing values of $\bar V$.

\section{Conclusions}
\label{section5}



\color{black}\subsection{Results}\color{black}


This paper analyses the influence of an external field on the stability (Hessian criterion and wrinkling bifurcation) of  an electro-elastic plate subject to an in-plane equi-biaxial contraction or extension. 
The plate is immersed in a tank filled with silicone oil with the electric field generated by two fixed rigid electrodes placed on  top and bottom of the tank. 
Provided the plate thickness is small compared to the lateral dimensions, which we assume to be the case, the electric field within the plate can be taken as uniform. 
We can then obtain the equations governing  the nonlinear response in terms of the non-dimensionalised  in-plane nominal stress $\bar s$  and the potential difference $\bar V$. 

We find that the response of the immersed plate depends  on the values of \color{black} $\bar \varepsilon =\varepsilon_{\mathrm d}/\varepsilon_{\mathrm f}$ \color{black}, the ratio of plate permittivity to that of the silicone oil, of \color{black} $H_{\mathrm d}/\mathcal L$\color{black}, the ratio of the plate thickness  to the \color{black} wavelength\color{black}, and of $\bar H=H_{\mathrm d}/H$, the ratio of the initial plate thickness to the distance of the electrodes. 
For example, an increase in the electric potential induces a thinning  of the plate and an accompanied lateral expansion when $\bar \varepsilon=0.1, 0.6$, similar to the behaviour in the absence of external field. 
But for the values  $\bar \varepsilon=1.5, 2.0$, the potential $\bar V$ generates an \textit{increase in plate thickness} and therefore a lateral contraction. 

First, we use the Hessian criterion to signal electro-mechanical instability.
When the electric field induces uniform thinning of the plate accompanied by a sudden increase in the in-plane stretch $\lambda$, we have the well-known \textit{pull-in instability}.
However, the Hessian criterion also identifies a \textit{pull-out instability}, when the plate thickens with the electric field.

Then we derive incremental forms of the governing equations and  boundary conditions in the Stroh form and obtain the explicit bifurcation conditions for antisymmetric and symmetric modes.  
We focus on a neo-Hookean electro-elastic material with a deformation-independent permittivity to illustrate the results. 


For the immersed plate, the curve showing the electric potential $\bar V$ versus the in-plane stretch $\lambda$, with pre-stress $\bar s=0$,   no longer represents the \color{black}wrinkling \color{black}  limit for plates with vanishing thickness. 
\color{black}Wrinkling \color{black}  in contraction or extension can now be delayed or suppressed by adjusting the values of the ratios $\bar H, \bar \varepsilon$ and  \color{black} $H_{\mathrm d}/\mathcal L$\color{black}.  The activation protocol analysed in this problem can therefore be used to optimise the design of electro-elastic sensors and actuators. 


\color{black}\subsection{Limitations}


In this paper, we consider a 2D incremental wrinkling deformation superposed upon a large equi-biaxial deformation. 
It is two-dimensional in the sense that the incremental fields depend on two space variables only, $x_1$ and $x_2$. 
It follows that on the plate surfaces at $x_2=\pm H_{\mathrm d}/2$ we see a 1D sinusoidal pattern emerge. 
In general, the deformed plate may buckle in either a one- or two-dimensional wrinkling pattern, depending on the material properties and loading conditions. 

From a practical viewpoint, we assumed that the incremental loading and boundary conditions superposed on the equi-biaxial deformed configuration cannot be exactly the same, and that the wrinkles' front is normal to the direction of the slightly larger load. Here we chose $x_1$ as the normal to the wrinkles' front, although we could have equally chosen any other direction in the ($x_1,x_3$) plane (and obtained the same critical thresholds). To predict a two-dimensional wrinkling pattern, a 3D incremental analysis is required, see Su et al.~\cite{Su2019b} for an example.
\color{black}

Note that  we focused on the \textit{static} \color{black}wrinkling \color{black}  behaviour of electro-elastic plates, and that viscosity \cite{Melcher1973} of the conductive fluid was not considered. 

\color{black}
We also studied the Hessian instability, but did not look at other instability modes such as, for example, the necking instability, see Fu et al. \cite{Fu2018}, and left aside the possibility of an electric breakdown of the DE plate.
For a detailed discussion on that latter topic, the interested readers are referred to recent papers by Su et al. \cite{Su2018,Su2018b}.
\color{black}

\enlargethispage{20pt}

\section*{Acknowledgments}
{This work was supported by a Government of Ireland Postdoctoral Fellowship from the Irish Research Council (No. GOIPD/2017/1208) and by the National Natural Science Foundation of China (No. 11621062).
WQC and YPS also acknowledge the support from the Shenzhen Scientific and Technological Fund for R$\&$D (No. JCYJ20170816172316775).}

{MD thanks Zhejiang University for funding research visits  to Hangzhou.}



\color{black}

\renewcommand\thesection{\Alph{section}}
\setcounter{section}{0}
\renewcommand*{\thesection}{A}


\section*{Appendix.  Stroh method for a neo-Hookean electro-elastic plate}


We restrict attention to an incompressible neo-Hookean electro-elastic plate with  underlying in-plane biaxial  deformation   $\lambda_1,\lambda_3$ and with  out-of-plane stretch $\lambda_2=\lambda_1^{-1}\lambda_3^{-1}$. The Stroh method is used to derive the bifurcation equation and details are provided.

Following  \cite{Su2018} and \cite{Dorfmann2019},  we consider an incompressible  neo-Hookean  electro-elastic material with the energy function
\begin{equation}
\label{neo-Hookean}
W(\lambda_1,\lambda_3,E_0)=\frac{\mu}{2}\left(\lambda_1^2+\lambda_3^2+\lambda_1^{-2}\lambda_3^{-2}-3\right)-\frac{\varepsilon_{\mathrm d}}{2}\lambda_1^{2}\lambda_3^{2}E_0^2,
\end{equation}
where $\mu$ is the shear modulus defined in the undeformed configuration and $\varepsilon_{\mathrm d}$ is the electric permittivity. Then, the explicit expressions of the coefficients in \eqref{stroh-components} are 
\begin{align}
& a=\mu\lambda_1^2\left(1-\lambda_3^2\bar  E_0^2\right),\quad b=\frac{\mu}{2}\left(\lambda_1^2+\lambda_1^{-2}\lambda_3^{-2}-\lambda_1^2\lambda_3^2\bar  E_0^2\right),\nonumber\\
&c=\mu\lambda_1^{-2}\lambda_3^{-2}, \quad d=\mu\lambda_1\lambda_3\bar  E_0, \quad e=2d,
\quad f=g=-\varepsilon_{\mathrm d},
\end{align}
where $\bar  E_0=E_0\sqrt{\varepsilon_{\mathrm d}/\mu}$ is the non-dimensional form of the nominal electric field in the plate.

\subsection*{(a) Static response}


The  nominal electric field component $E_0$ is obtained from Eq.~\eqref{electric-field} using the energy function  \eqref{neo-Hookean}.  This component, in non-dimensional form, has the expression
\begin{equation}
\label{VE2}
\bar  E_0=\frac{\bar  V}{\bar  H\left(1-\bar  \varepsilon\right)+\bar  \varepsilon\lambda_1\lambda_3},
\end{equation}
where \color{black} $\bar  V=V\sqrt{\varepsilon_{\mathrm d}/\mu}/H$  is the non-dimensionalized form of the electric potential, \color{black} $\bar  H=H_{\mathrm d}/H$ is  the ratio of the initial plate thickness  to the distance of the electrodes and  $\bar  \varepsilon=\varepsilon_{\mathrm d}/\varepsilon_{\mathrm f}$ denotes the ratio  of the plate permittivity to that of the silicone oil. 

For the neo-Hookean electro-elastic material \eqref{neo-Hookean}, the  expressions of the  in-plane nominal stress components $s_1,s_3$ are obtained from \eqref{exterior-stressa} and  \eqref{problem2}. These are, in non-dimensionalised form
\begin{equation}\label{nonlinear-response-biaxial-2}
\bar  s_1=\lambda_1-\lambda_1^{-3}\lambda_3^{-2}-\left(1-\bar \varepsilon\right)\bar  E_0^2\lambda_1\lambda_3^2, \quad
\bar  s_3=\lambda_3-\lambda_1^{-2}\lambda_3^{-3}-\left(1-\bar \varepsilon\right)\bar  E_0^2\lambda_1^2\lambda_3.
\end{equation}
where $\bar  s_i=s_i/\mu \ (i=1,3)$.


\subsection*{(b) Wrinkling stability analysis}


The eigenvalues are obtained by solving  Eq. \eqref{eigensystem}, which  can be written in  compact form as
\begin{align}
p_1=-p_4=\lambda_1^2\lambda_3, \qquad p_2=-p_5=p_3=-p_6=1.
\end{align}
The component forms of the corresponding eigenvectors $\boldsymbol \eta^{(i)}, i=1, \dots,6,$ are
\begin{align}
&\boldsymbol{\eta}^{(1)}=\left[\begin{matrix}
&2\textrm{i} \lambda_1^4\lambda_3^3 \\[3pt]
&-2\lambda_1^{2}\lambda_3^{2} \\[3pt]
&2\textrm{i} \sqrt{\varepsilon_{\mathrm d}\mu} \lambda_1^{5}\lambda_3^{4}\bar  E_0 \\[3pt]
&\mu\lambda_1^{4}\lambda_3^{2}\left[\lambda_3^2\bar  E_0^2\left(\bar \varepsilon-2\right)-2\right] -2\mu\\[3pt]
&\textrm{i}\mu\lambda_1^{2}\lambda_3\left(\lambda_1^{4}\lambda_3^{4}\bar \varepsilon\bar  E_0^2-4\right)\\[3pt]
&2\sqrt{\mu/\varepsilon_{\mathrm d}}\lambda_1^{3}\lambda_3^{3}\bar  E_0
\end{matrix}\right], \nonumber \\[4pt]
&\boldsymbol{\eta}^{(2)}=\left[\begin{matrix}
&2\textrm{i} \lambda_1^2\lambda_3^3 \\[3pt]
&-2\lambda_1^{2}\lambda_3^{2} \\[3pt]
&4\textrm{i}\sqrt{\varepsilon_{\mathrm d}\mu}\lambda_1^{3}\lambda_3^{3}\bar  E_0 \\[3pt]
&\mu\left(\lambda_1^{4}\lambda_3^{4}\bar \varepsilon\bar  E_0^2-4\right) \\[3pt]
&\textrm{i}\mu\lambda_1^{4}\lambda_3^{2}\left[\lambda_3^2\bar  E_0^2\left(\bar \varepsilon+2\right)-2\right]-2\mu\\[3pt]
&0
\end{matrix}\right],\nonumber\\[6pt]
&\boldsymbol{\eta}^{(3)}=\left[\begin{matrix}
&2\textrm{i} \lambda_1^3\lambda_3^3\bar  E_0 \\[3pt]
&2\lambda_1^3\lambda_3^3\bar  E_0 \\[3pt]
&\textrm{i} \sqrt{\varepsilon_{\mathrm d}\mu} \lambda_1^{4}\lambda_3^{2}\left[2-\lambda_3^2\bar  E_0^2\left(\bar \varepsilon-2\right)\right]+2\textrm{i} \sqrt{\varepsilon_{\mathrm d}\mu}  \\[3pt]
&-2\mu\lambda_1\lambda_3\bar  E_0\left(\lambda_1^{4}\lambda_3^4\bar  E_0^2-\lambda_1^{4}\lambda_3^2+1\right) \\[3pt]
&0\\[3pt]
&\sqrt{\varepsilon_{\mathrm d}\mu} \lambda_1^{4}\lambda_3^{2}\left[\lambda_3^2\bar  E_0^2\left(\bar \varepsilon+2\right)-2\right]-2\sqrt{\varepsilon_{\mathrm d}\mu}
\end{matrix}\right], 
\nonumber \\[6pt]
& \boldsymbol{\eta}^{(4)}=\left( {\boldsymbol{\eta}}^{(1)}\right)^*, \qquad 
   \boldsymbol{\eta}^{(5)}=\left( {\boldsymbol{\eta}}^{(2)}\right)^*, \qquad 
   \boldsymbol{\eta}^{(6)}=\left( {\boldsymbol{\eta}}^{(3)}\right)^*.
\end{align}

The \color{black}wrinkling \color{black}  criterion for antisymmetric modes is obtained from \eqref{asymmetric2} and has the explicit non-dimensional form
\color{black}
\begin{align}
\label{bifurcation2}
\bar  E_0=&\sqrt{\bar \varepsilon\tanh\left[\pi \left(\bar H-\lambda_1^{-1}\lambda_3^{-1}\right)H_{\mathrm d}/{\mathcal L}\right]+\coth\left(\pi \lambda_1^{-1}\lambda_3^{-1} H_{\mathrm d}/\mathcal L\right)}  \nonumber\\[4pt]
& \quad \sqrt{\frac{\left(\lambda_1^4\lambda_3^2+1\right)^2\tanh\left(\pi \lambda_1^{-1}\lambda_3^{-1}H_{\mathrm d}/\mathcal L\right)-4\lambda_1^2\lambda_3\tanh\left(\pi \lambda_1 H_{\mathrm d}/\mathcal L\right)}{\lambda_1^4\lambda_3^4\left(\bar \varepsilon-1\right)^2\left(\lambda_1^4\lambda_3^2-1\right)}}.
\end{align} 
\color{black}
The thin- and thick-plate \color{black}wrinkling \color{black}  \color{black} criteria \color{black}are obtained by evaluating \eqref{bifurcation2}   for $H/\mathcal L \to 0$   and $H/\mathcal L \to \infty$, respectively. In non-dimensional forms, these are given by
\begin{equation}
\bar  E_0=\sqrt{\frac{\lambda_1^{4}\lambda_3^2-1}{\lambda_1^{4}\lambda_3^4\left(\bar \varepsilon-1\right)^2}}
\end{equation}
and
\begin{equation}
\bar  E_0=\sqrt{\frac{\left(\bar \varepsilon+1\right)\left[\lambda_1^2\lambda_3\left(\lambda_1^4\lambda_3^2+\lambda_1^2\lambda_3+3\right)-1\right]}{\lambda_1^4\lambda_3^4\left(\bar \varepsilon-1\right)^2\left(\lambda_1^2\lambda_3+1\right)}}.
\end{equation}

Eq.~\eqref{bifurcation2}, using the connection \eqref{VE2} with $\bar V=0$, identifies a configuration where an  antisymmetric  wrinkling mode may exist for a purely elastic plate. 


\end{document}